\documentclass[conference]{IEEEtran}
\IEEEoverridecommandlockouts
\usepackage{cite}
\usepackage{amsmath,amssymb,amsfonts}
\usepackage{algorithmic}
\usepackage{graphicx}
\usepackage{textcomp}
\usepackage{xcolor}
\usepackage{threeparttable}
\usepackage{booktabs,multirow}
\usepackage{amsfonts}
\def\BibTeX{{\rm B\kern-.05em{\sc i\kern-.025em b}\kern-.08em
    T\kern-.1667em\lower.7ex\hbox{E}\kern-.125emX}}
\begin{document}

\title{SMC-ITA: Sequential Monte Carlo Inference-Time Alignment for Video-to-Audio Generation}

\author{
\IEEEauthorblockN{
Haoyu Zhang\IEEEauthorrefmark{1},
Yuta Oshima\IEEEauthorrefmark{1},
Xingjian Du\IEEEauthorrefmark{2},
Chunfeng Wang\IEEEauthorrefmark{3},
Irene Li\IEEEauthorrefmark{1},
Yusuke Iwasawa\IEEEauthorrefmark{1},
and Yutaka Matsuo\IEEEauthorrefmark{1}
}

\IEEEauthorblockN{
\IEEEauthorrefmark{1}
The University of Tokyo, Tokyo, Japan\\
E-mail: haoyu.zhang@weblab.t.u-tokyo.ac.jp}

\IEEEauthorblockN{
\IEEEauthorrefmark{2}
University of Rochester, New York, USA}

\IEEEauthorblockN{
\IEEEauthorrefmark{3}
Independent, Beijing, China}
}

\maketitle

\begin{abstract}
Video-to-audio (V2A) generation must jointly satisfy audiovisual alignment, semantic consistency, temporal synchronization, and perceptual quality. While prior work has mainly focused on model architecture, multimodal conditioning, and training objectives, inference-time alignment for V2A remains underexplored. In this paper, we study inference-time alignment for flow-matching-based V2A generation and formulate it as a search problem. We propose Sequential Monte Carlo Inference-Time Alignment (SMC-ITA), which combines lookahead-based reward estimation and sequential Monte Carlo resampling to reallocate computation adaptively using multi-dimensional cross-modal rewards. SMC-ITA improves over naive single-trajectory sampling, achieving a 55.67\% relative reduction in DeSync, a 20.23\% improvement in IB-score, and a 15.44\% improvement in Audio Quality. Under matched NFE budgets, it also achieves the best overall trade-off among the compared search baselines, outperforming Best-of-N and Beam Search. Ablation studies further show that lookahead improves the reliability of intermediate reward estimates and that systematic resampling is a strong practical default for V2A inference-time alignment.
\end{abstract}

\section{Introduction}
Video-to-audio (V2A) aims to generate audio for silent videos, optionally conditioned on text. Successful V2A generation must satisfy not only perceptual quality but also semantic consistency, audiovisual alignment, temporal synchronization \cite{luo2023difffoley,zhang2026foleycrafter,vatt,cheng2025mmaudio,liu2026prismaudio,simon2026echoes,fang2026acfoley}. In practice, these requirements are difficult to capture in a single objective, and some are easier to assess using multiple reward functions. This motivates reward-guided inference-time search for V2A generation. However, compared with efforts in image, video, and text-to-audio generation, reward-guided inference-time search remains less explored for V2A~\cite{lee2023aligning,kim2025testtime,huang2024symbolic,oshima2025inference,jung2025score}.
In particular, the intermediate steps of flow matching often yield noisy rewards in the early stages, when the search space is still large.
This makes inference-time search both challenging and beneficial.

To address this issue, we propose Sequential Monte Carlo Inference-Time Alignment (SMC-ITA), an inference-time alignment method for V2A generation that aligns the generation trajectory with the input conditions by optimizing multi-dimensional rewards. Specifically, SMC-ITA employs Sequential Monte Carlo sampling~\cite{del2006sequential}, also known as particle filtering, to steer generation toward high-reward trajectories while preserving other promising candidates. In addition, the proposed lookahead strategy mitigates reward noise in the early stages, enabling a more accurate and effective search.
Experiments show that SMC-ITA substantially improves generation over naive single-trajectory sampling, yielding a 55.67\% relative reduction in DeSync, a 20.23\% improvement in IB-score, and a 15.44\% improvement in Audio Quality. Under the same NFE budget, it also achieves the strongest overall results among the compared search baselines.

In summary, our contributions are: 1) we study inference-time alignment for V2A generation as a search problem with multi-dimensional cross-modal rewards; 2) we propose SMC-ITA, an inference-time alignment method that combines lookahead-based reward estimation with sequential Monte Carlo resampling to support effective search under noisy intermediate rewards; and 3) we show that inference-time search improves over naive sampling, while SMC-ITA outperforms Best-of-N and Beam Search under matched NFE budgets.

\section{Related Work}
{\bf Video-to-audio generation}. 
Previous V2A work has focused primarily on improving model architecture, multimodal conditioning, and training objectives. 
Diff-Foley\cite{luo2023difffoley} aligns video and audio with contrastive learning and uses aligned embedding to guide audio generation. Foleycrafter~\cite{zhang2026foleycrafter} improves multimodal alignment from a pretrained text-to-audio generation model through semantic and temporal adaptation. VATT\cite{vatt} employs a two-stage training framework that first aligns video and text with a large language model and then learns audio generation. MMAudio~\cite{cheng2025mmaudio} jointly trains multimodal and audio-only transformers and improves audio-visual synchronization with Synchformer features~\cite{iashin2024synchformer}. PrismAudio~\cite{liu2026prismaudio} uses a multimodal video-language model to generate decomposed CoTs and improves V2A generation through Fast-GRPO with multi-dimensional rewards. MMHNet~\cite{simon2026echoes} extends the generation of V2A with an HNet architecture~\cite{hwang2026dynamic} and improves the generalization of the length for the generation of long videos to audio. AC-Foley~\cite{fang2026acfoley} introduces a video-to-audio generation framework with audio conditioning. Overall, existing V2A methods mainly improve semantic and temporal alignment through architectural and training advances, while inference-time alignment remains largely unexplored.

{\bf Inference-time alignment in diffusion models}. A growing body of work studies inference-time steering/alignment for diffusion and flow-based generative models, often without updating model parameters~\cite{ma2025inference,singhal2025a}. Dhariwal and Nichol~\cite{dhariwal2021diffusion} propose classifier guidance to steer diffusion sampling by adding classifier gradients to the score estimate.
DLBS~\cite{oshima2025inference} performs inference-time beam search over diffusion trajectories to improve text-to-video alignment.
DSearch~\cite{li2025dynamic} formulates inference-time alignment in diffusion models as a search problem and dynamically adjusts tree expansion to optimize non-differentiable rewards efficiently. DAS~\cite{kim2025testtime} proposes a training-free inference-time alignment method based on SMC, which steers diffusion sampling toward reward-aligned trajectories.
EvoSearch~\cite{he2025scaling} performs inference-time evolutionary search over diffusion/flow denoising trajectories via selection and mutation to improve image and video generation. SCORE~\cite{jung2025score} extends inference-time scaling to text-to-audio generation and proposes a composite multi-reward guidance that improves both perceptual quality and text–audio alignment. Inference-time alignment is particularly challenging for V2A generation, where the model must simultaneously satisfy consistency constraints across text, video, and audio.

\section{Preliminaries}
This section presents conditional flow matching and its stochastic differential equation (SDE) solver.

\subsection{Conditional flow matching}

We conduct our inference-time experiments using a model trained with the conditional flow matching (CFM) objective~\cite{lipman2023flow}. Let $x_0 \sim X_0$ denote a sample from the standard normal distribution and $x_1 \sim X_1$ denote a sample from the data distribution. The model is trained to estimate the velocity field by minimizing
\begin{equation}
    \mathbb{E}_{t, x_0 \sim X_0, x_1 \sim X_1} \|v_\theta(x_t, t \mid c) - v(x_t|x_0, x_1)\|^2,
\end{equation}
where $t \in [0,1]$ and $c$ denotes the conditioning information, which in our setting consists of video and text. The intermediate state is given by $x_t = (1-t)x_0 + tx_1$ and the corresponding target velocity is $v(x_t \mid x_0, x_1) = x_1 - x_0$.

At inference time, sample generation is performed by integrating the learned velocity field with an ODE solver. Using Euler discretization, the state is updated as
\begin{equation}
x_{t+\Delta t} = x_t + \Delta t \, v_\theta(x_t, t, \mid c).
\label{eq:ode}
\end{equation}
Hereafter, we omit $c$ when clear from context.

\subsection{Extend ODE to SDE}
To enable beam search or SMC resampling in flow matching, the sampling process must provide stepwise stochasticity, allowing multiple candidate trajectories to be propagated and evaluated. In standard flow matching, however, sampling is performed by a deterministic ODE solver, which can only generate a single trajectory from the initial normal distribution and cannot introduce randomness in later steps. To address this issue, we instead apply the SDE solver. In our setting, the direction of the probability path follows the flow matching convention, transforming noise into data. Following ~\cite{song2020score}, we construct the SDE formulation:
\begin{equation}
    dx_t =
\left(
v_t(x_t)
+\frac{\sigma_t^2}{2}\nabla \log p_t(x_t)
\right) dt+ \sigma_t dw,
\end{equation}
where $dw$ denotes the increment of a standard Wiener process and $\sigma_t$ is the diffusion coefficient.
Applying Euler-Maruyama discretization yields the following update rule:
\begin{equation}
\begin{split}
x_{t+\Delta t}=&x_t+v_{drift}(x_t,t)\Delta t+\sigma_t\sqrt{\Delta t}\,\epsilon, \\
v_{drift}(x_t,t)=&
v_\theta(x_t,t)-\frac{\sigma_t^2}{2(1-t)}
\left(x_t - t\,v_\theta(x_t,t)\right),
\end{split}
\end{equation}
where $\epsilon \sim \mathcal{N}(0, \mathcal{I})$ is Gaussian noise and $\sigma_t$ controls the step-wise noise level. We set $\sigma_t = \sigma_{\text{base}} \cos\left(\frac{\pi t}{2}\right)$ in experiments.

\section{SMC-ITA}

\begin{figure*}[htbp]
\centering
\includegraphics[width=0.75\textwidth]{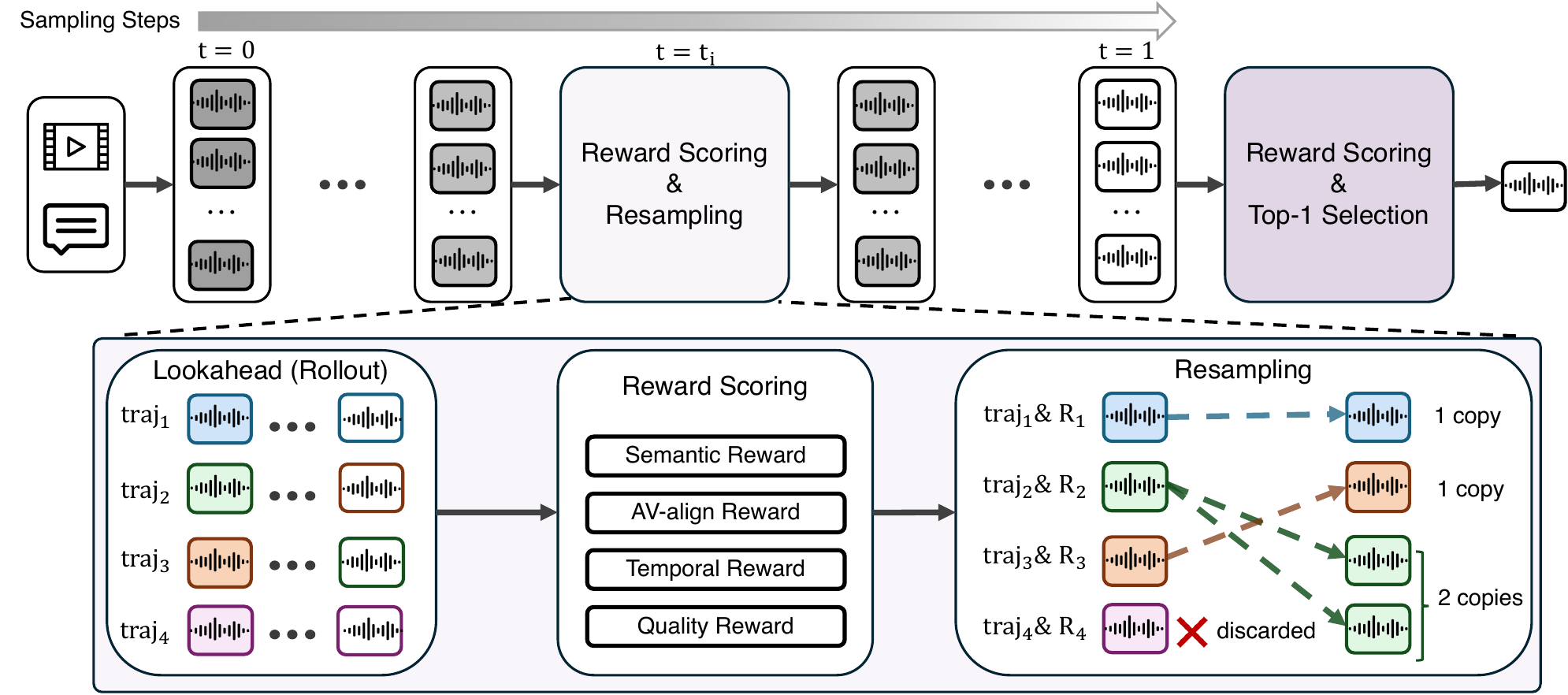}
\caption{Overview of SMC-ITA. $traj_k$ denotes the $k$-th trajectory, and $R_k$ denotes its aggregated reward. At each search step, lookahead performs a fast rollout from each trajectory to the clean audio samples, after which the reward is computed from the rollout results. Resampling then replicates high-reward trajectories and discards low-reward ones.}

\label{fig:frame}
\end{figure*}

\subsection{Overview}
\label{iv:overview}
We study Inference-Time Alignment (ITA) for video-to-audio generation. Given video and text conditions, the model generates audio by guiding the flow matching sampling process. A natural baseline is naive sampling, which follows a single trajectory throughout the generation process. A stronger baseline is Best-of-N (BoN), which runs multiple trajectories in parallel and selects the best sample only after the final step. Beam search provides another competitive baseline by scoring trajectories and pruning them at predefined search steps. However, such hard selection can prematurely discard trajectories that appear suboptimal at early stages but may become promising later, thereby limiting overall performance. In contrast, our method performs search during sampling in a softer and more adaptive manner. As shown in Fig.~\ref{fig:frame}, we maintain a population of candidate trajectories and intervene at a set of predefined search steps, $T_{\mathrm{search}}=\{t_1,\ldots,t_i\}$. At each search step, we evaluate the current trajectories using reward functions, convert these scores into weights, and resample the population accordingly. As a result, trajectories with stronger intermediate alignment signals are more likely to survive and proliferate, while weaker ones are gradually removed. This sequential weighting-and-resampling scheme progressively allocates computation to promising regions of trajectory space, rather than committing to a single path or postponing selection until the end. To make such a search effective, trajectory diversity is essential. We therefore adopt an SDE solver, which introduces stochasticity into flow matching and prevents different trajectories from collapsing to nearly identical paths. This stochasticity enlarges the effective search space and enables inference-time search to improve alignment.

\subsection{Reward Scoring}
For video-to-audio generation, alignment must be established across three modalities: video, text, and audio. To better capture different aspects of high-quality generation, we define four reward functions: 1) Semantic Reward measures text-audio consistency using the cosine similarity between the input text and the generated audio, computed with LAION-CLAP~\cite{htsatke2022,laionclap2023}; 2) AV-align Reward measures video-audio consistency using the cosine similarity between the input video and the generated audio, computed with ImageBind~\cite{girdhar2023imagebind}; 3) Temporal Reward measures temporal synchronization between video and audio using Synchformer~\cite{iashin2024synchformer};
4) Quality Reward measures the perceptual quality of the generated audio using the Product Quality score from Audiobox-Aesthetics~\cite{tjandra2025aes}.

Since these rewards differ in scale and distribution, we z-normalize each reward and define the aggregated reward of a predicted sample $\hat x$ as
\begin{equation}
R(\hat x, c)=\frac{1}{|\mathcal{M}|}\sum_{m\in\mathcal{M}}\frac{r_m(\hat x,c)-\mu_m}{\sigma_m},
\end{equation}
where $\mathcal{M}=\{\mathrm{Semantic},\,\mathrm{AV\text{-}Align},\,\mathrm{Temporal},\,\mathrm{Quality}\}$. 
Here, $r_m(\cdot)$ denotes the $m$-th reward function. $\mu_m$ and $\sigma_m$ are the corresponding mean and standard deviation estimated on the VGGSound validation set. We use the same reward configuration in all experiments for fair comparison.

\subsection{Lookahead and Resampling Strategy}
In flow matching models, the early stages offer a larger search space, but the reward signal is often noisy, whereas later stages provide more reliable rewards but much less room for search. Therefore, effective early-stage search is crucial for strong final performance. To address this challenge, we adopt a lookahead strategy for better reward estimation:
\begin{equation}
\hat{x}_{1|t_i}=\mathrm{LA}_{t_i \rightarrow 1}^{(l_a)}(x_{t_i}),
\end{equation}
where $\mathrm{LA}^{(l_a)}_{t_i\rightarrow 1}(\cdot)$ denotes an $l_a$-step rollout from $t_i$ to $1$ using the sampler in Eq.~\eqref{eq:ode}, with the remaining interval $[t_i,1]$ uniformly divided into $l_a$ substeps of size $\Delta_i=(1-t_i)/l_a$. We then use $R(\hat{x}_{1|t_i}, c)$ to score each trajectory at step $t_i$. This helps retain trajectories that are more promising in the long run and reduces the impact of noisy intermediate rewards. To keep the overall NFE unchanged, we reduce the population size $N$ accordingly.

At each search step $t_i$, we compute a normalized importance weight $w$ for each trajectory $n$:
\begin{equation}
w_{t_i}^{(n)}=
\frac{\exp\!\left(\tau\, R(\hat{x}_{1|t_i}^{(n)}, c)\right)}
{\sum_{\ell=1}^{N}\exp\!\left(\tau\, R(\hat{x}_{1|t_i}^{(\ell)}, c)\right)},
\end{equation}
where $\tau$ serves as a temperature parameter controlling the sharpness of the resampling distribution.
Then we perform resampling using these weights.
We choose systematic resampling~\cite{carpenter1999improved} as the default because it is low-overhead and empirically effective.
Specifically, we draw a single random offset and select $N$ trajectories using evenly spaced samples on the cumulative weight distribution.
We also evaluate alternative resampling schemes, including Srinivasan Sampling Process resampling~\cite{gerber2019negative} in our ablation. 
Beyond these SMC resampling variants, we also compare with EvoSearch~\cite{he2025scaling} as an additional population-based resampling variant.

\begin{table*}[t]
\begin{center}
\begin{threeparttable}
\caption{Video-to-audio results on the VGGSound test set 1k.}
\small
\setlength{\tabcolsep}{4pt}
\renewcommand{\arraystretch}{1.15}
\begin{tabular}{l  c c c c c c  c c  c c c}
\toprule
\multirow{2}{*}{Methods} & \multicolumn{5}{c}{Generation quality}
& \multicolumn{2}{c}{AV align.}
& \multicolumn{1}{c}{Temporal align.}
& \multicolumn{3}{c}{Audiobox-Aesthetics} \\
\cmidrule(lr){2-6}
\cmidrule(lr){7-8}
\cmidrule(lr){9-9}
\cmidrule(lr){10-12}
& FD\textsubscript{PANNs}$\downarrow$ & 
FD\textsubscript{PaSST}$\downarrow$ & KL\textsubscript{PANNs}$\downarrow$ & 
KL\textsubscript{PaSST}$\downarrow$  & AQ$\uparrow$ \tnote{a} & IB-score$\uparrow$ & AVCC$\uparrow$\tnote{a} & DeSync$\downarrow$ & PQ$\uparrow$\tnote{b} & CU$\uparrow$\tnote{b} & CE$\uparrow$\tnote{b} \\
\midrule

Naive Sampling & \textbf{12.274} & 144.983 & 1.646 & 1.657 & 18.854 & 28.374 & 18.684 &0.494 & 5.721 & 5.137 & 3.878\\
\midrule
Best-of-N& 12.998 & 145.750 & 1.590 & 1.585 & 21.106 & 32.740 & 20.702 & 0.254 & 6.031 & 5.459 & 4.060\\
\midrule
Beam Search & 12.876 & \textbf{144.489} & 1.602 & 1.592 & 21.080 & 33.201 & 20.994 & 0.252 & 6.064 & 5.475 & 4.080\\
\midrule
SMC-ITA (Ours) & 12.932 & 144.948 & \textbf{1.578} & \textbf{1.578} & \textbf{21.764} & \textbf{34.114} & \textbf{21.666} & \textbf{0.219} & \textbf{6.126} & \textbf{5.530} & \textbf{4.103}\\

\bottomrule
\end{tabular}
\begin{tablenotes}
\item[a] AGAV-Rater scores: audio quality (AQ) and audio-visual content consistency (AVCC).
\item[b] Audiobox-Aesthetics scores: production quality (PQ), content usefulness (CU), and content enjoyment (CE).
\end{tablenotes}
\label{tab:main}
\end{threeparttable}
\end{center}
\vspace{-2mm}
\end{table*}

\section{Experiments}
In this section, we introduce our experiment setting and the performance on different methods.

\subsection{Experiment Settings}
We evaluate our method on a 1k subset of the VGGSound test set~\cite{chen2020vggsound}. MMAudio-S-16kHz~\cite{cheng2025mmaudio} serves as the base model in all inference-time alignment experiments. We use AV-Benchmark~\cite{cheng2025mmaudio}, Audiobox-Aesthetics~\cite{tjandra2025aes}, and AGAV-Rater~\cite{cao2025agavrater} for evaluation.

We evaluate the four strategies introduced in Sec.~\ref{iv:overview}: Naive Sampling, Best-of-N (BoN), Beam Search, and SMC-ITA. Beam Search follows a DLBS-style~\cite{oshima2025inference} expansion and pruning procedure adapted to V2A flow matching and uses lookahead-based reward estimation. Beam Search and SMC-ITA use the same predefined search steps. For the main experiment, except for Naive Sampling, all methods use the same reward-scoring configuration and are compared under the same computational budget (NFE = 800). 
 BoN uses 16 parallel trajectories; Beam Search uses beam width 4, expansion factor 5, and lookahead $l_a=3$; SMC-ITA uses population size 10, $l_a = 3$, systematic resampling, and temperature $\tau=10$. 
All experiments are conducted on a single GH200 GPU.

\subsection{Main Results}
\label{ssec:mainexp}
Table~\ref{tab:main} presents the main results. Compared with Naive Sampling, the inference-time methods improve most metrics, while showing mixed behavior on FD-based generation-quality metrics, suggesting that the proposed reward formulation is effective. Beam Search is a strong baseline, but its gains over BoN are limited. In contrast, SMC-ITA achieves the best overall trade-off among the compared methods and outperforms Beam Search. Although it does not obtain the best FD score, it performs best on KL, AQ, IB-score, AVCC, DeSync, PQ, CU, and CE, showing consistent advantages across generation quality, alignment, and perceptual evaluation. 
This advantage is consistent with the benefits of lookahead-based reward estimation and resampling.
Compared with BoN, SMC-ITA can reallocate computation during sampling rather than selecting only at the end; compared with Beam Search, it preserves promising candidates without aggressively discarding alternatives. Finally, consistent gains on AQ and AVCC from AGAV-Rater indicate that the improvements are not confined to the internal reward functions used during search.

\subsection{Ablation Studies}
\label{ssec:ablation}

\begin{table}[htbp]
\vspace{-1.5mm}
\caption{Rewards ablation.}
\begin{center}
\setlength{\tabcolsep}{3pt} 
\begin{tabular}{lccccc}
\toprule
Method & KL\textsubscript{PANNs}$\downarrow$ & AQ $\uparrow$& IB-score$\uparrow$ & DeSync$\downarrow$ & PQ$\uparrow$\\
\midrule
Naive Sampling & 1.646 & 18.854 & 28.374 & 0.494 & 5.721 \\
\midrule
\multicolumn{6}{@{}l}{\textbf{\emph{Ours}}} \\
+ Semantic Reward & 1.586 & 19.955 & 28.785 & 0.511 & 5.753 \\
+ AV-align Reward& 1.597 & 20.621 & \textbf{37.243} & 0.491 & 5.739 \\
+ Temporal Reward & 1.628 & 19.058 & 28.524 & \textbf{0.135} & 5.730 \\
+ Quality Reward & 1.678 & 20.633 & 28.415 & 0.491 & \textbf{6.430}\\
+ All Rewards & \textbf{1.578} & \textbf{21.764} & 34.114 & 0.219 & 6.126 \\

\bottomrule
\end{tabular}
\label{tab:reward}
\end{center}
\vspace{-0.5mm}
\end{table}

\textbf{Effect of Rewards.} In the final system, we use four reward functions to steer audio generation. To verify whether each reward contributes as intended, we perform ITA experiments in which each reward is applied individually. As reported in Table~\ref{tab:reward}, we observe that the AV-alignment reward mainly improves IB-score, the temporal reward improves DeSync, and the quality reward improves PQ. The semantic reward primarily models text-audio consistency; although it is not directly aligned with the video-to-audio objective, it still provides consistent overall improvements. This demonstrates that SMC-ITA can effectively accommodate different reward designs, making the framework adaptable to different scenarios. When all rewards are combined, the model achieves the best overall trade-off across evaluation dimensions, although single-reward variants remain strongest on their corresponding target metrics.

\begin{table}[htbp]
\vspace{-0.5mm}
\caption{Effect of resampling schemes.}
\begin{center}
\setlength{\tabcolsep}{3pt} 
\begin{tabular}{lccccc}
\toprule
Resample schemes & KL\textsubscript{PANNs}$\downarrow$ & AQ $\uparrow$& IB-score$\uparrow$ & DeSync$\downarrow$ & PQ$\uparrow$\\
\midrule
EvoSearch & 1.605 & 21.757 & 33.688 & 0.224 & 6.108 \\
\midrule
SSP & 1.595 & 21.761 & 33.941 & \textbf{0.216} & \textbf{6.131}\\
\midrule
Systematic (Selected) & \textbf{1.578} & \textbf{21.764} & \textbf{34.114} & 0.219 & 6.126 \\

\bottomrule
\end{tabular}
\label{tab:remsample}
\end{center}
\vspace{-0.5mm}
\end{table}

\textbf{Effect of resampling schemes.} SMC-ITA introduces an adaptive resampling framework for flow matching generation, in which the resampling strategy plays a central role. We evaluate three alternatives, namely systematic resampling, Srinivasan Sampling Process (SSP) resampling, and EvoSearch. As reported in Table~\ref{tab:remsample}, both SSP and systematic resampling perform well, while systematic resampling achieves the best overall trade-off across different evaluation dimensions. Therefore, we choose systematic resampling in the final system.

\begin{table}[h]
\vspace{-0.5mm}
\caption{Effect of Lookahead.}
\begin{center}
\setlength{\tabcolsep}{3pt} 
\begin{tabular}{lccccc}
\toprule
Lookahead steps  & KL\textsubscript{PANNs}$\downarrow$ & AQ $\uparrow$& IB-score$\uparrow$ & DeSync$\downarrow$ & PQ$\uparrow$\\
\midrule
$l_a=0$ & 1.635 & 20.686 & 31.519 & 0.232 & 5.942 \\
\midrule
$l_a=1$ & 1.638 & 21.189 & 32.563 & 0.246 & 6.079\\
\midrule
$l_a=3$ (Selected) & \textbf{1.578} & \textbf{21.764} & \textbf{34.114} & 0.219 & \textbf{6.126} \\
\midrule
$l_a=5$  & 1.618 & 21.195 & 33.814 & \textbf{0.211} & 6.124 \\

\bottomrule
\end{tabular}
\label{tab:la}
\end{center}
\vspace{-0.5mm}
\end{table}

{\bf Effect of Lookahead.} To obtain more reliable reward estimates, we incorporate lookahead into SMC-ITA. Table~\ref{tab:la} shows that lookahead improves overall performance over the no-lookahead setting. However, under a fixed NFE budget, a larger lookahead horizon reduces the parallel search budget. Since $l_a=3$ performs substantially better than $l_a=1$ and slightly better than $l_a=5$, we use it in the final system as the practical trade-off between parallelism and lookahead.

\subsection{Further Analysis}
\label{ssec:anal}
\begin{figure}[t]
\centering
\includegraphics[width=0.95\columnwidth]{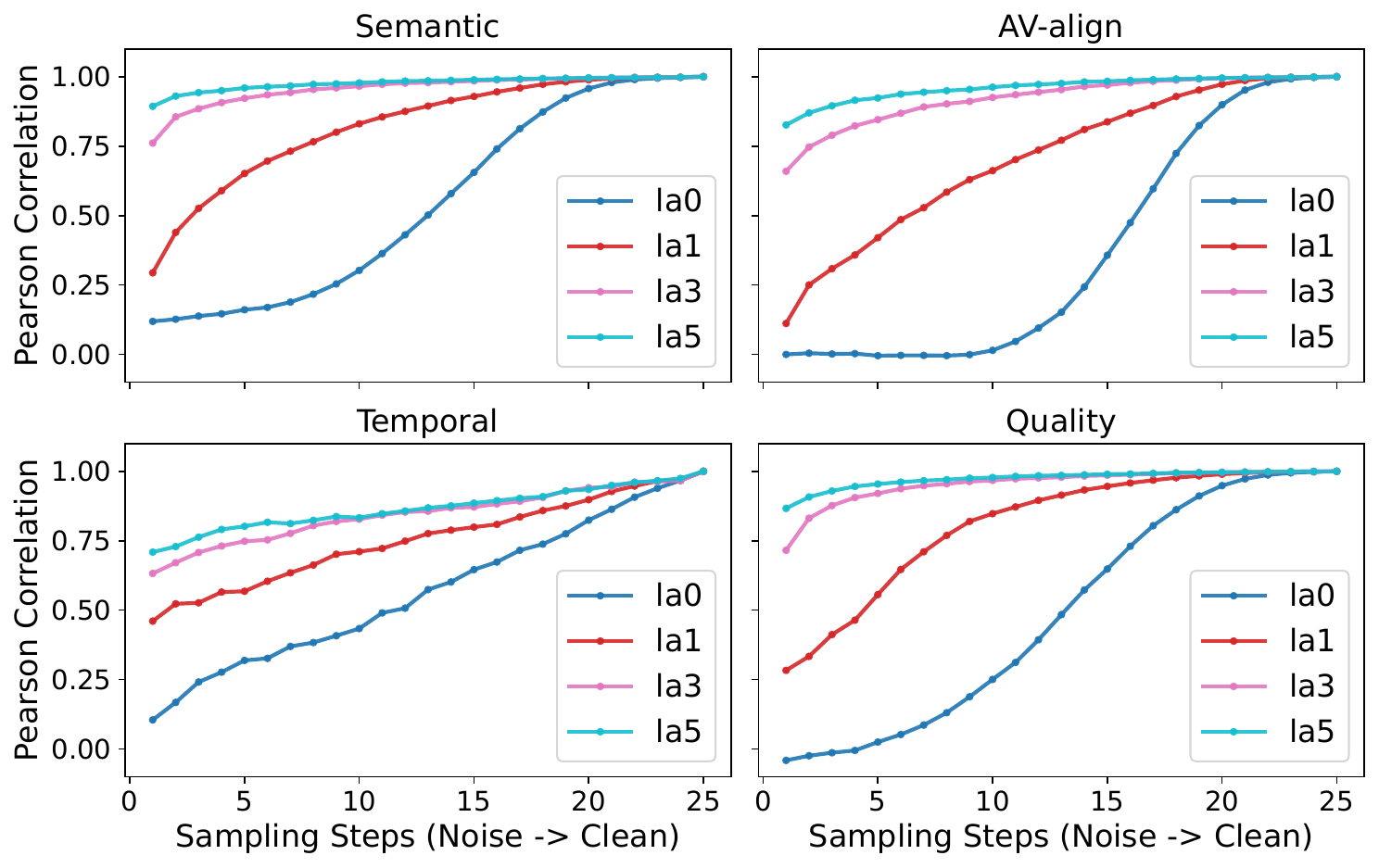}
\caption{Correlation between the reward at each intermediate step and the final reward. We observe that the correlation increases with the number of steps. Furthermore, increasing the lookahead makes the estimated rewards at intermediate steps more accurate predictors of the final reward.}
\label{fig:la}
\vspace{-0.5mm}
\end{figure}

{\bf Reward correlation between intermediate and final steps.}
Accurate search under a large search space is critical for successful inference-time alignment. As shown in Fig.~\ref{fig:la}, we measure the correlation between the reward at each intermediate step and the final-step reward. When the lookahead step is zero, early-stage rewards show low correlation with the final reward, indicating that they are not sufficiently reliable for effective search. As the generation proceeds, the trajectory becomes increasingly deterministic, leaving less room for further search. Introducing lookahead helps address this issue by making intermediate-step rewards more predictive of the final reward.

\begin{figure}[t]
\centering
\includegraphics[width=1.0\columnwidth]{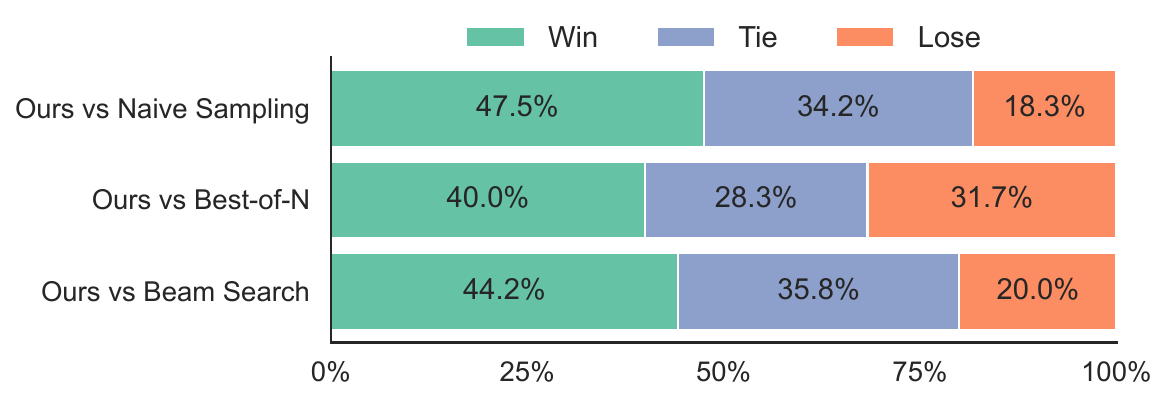}
\caption{Pairwise human evaluation against the baselines.}
\label{fig:ab}
\vspace{-1mm}
\end{figure}

{\bf Human Evaluation.}
To provide complementary subjective evidence, we conduct a pairwise human preference study on 20 test samples. 
We compare our method against Naive Sampling, Best-of-N, and Beam Search. 
Each pair is evaluated by six human raters, who select the preferred result or a tie based on overall audiovisual quality. 
Fig.~\ref{fig:ab} summarizes the aggregated win/tie/lose rates, showing that our method is preferred over all three baselines. These results are consistent with the objective metrics and indicate that the improvements of SMC-ITA are perceptible to human raters.

\begin{figure}[t]
\centering
\includegraphics[width=0.92\columnwidth]{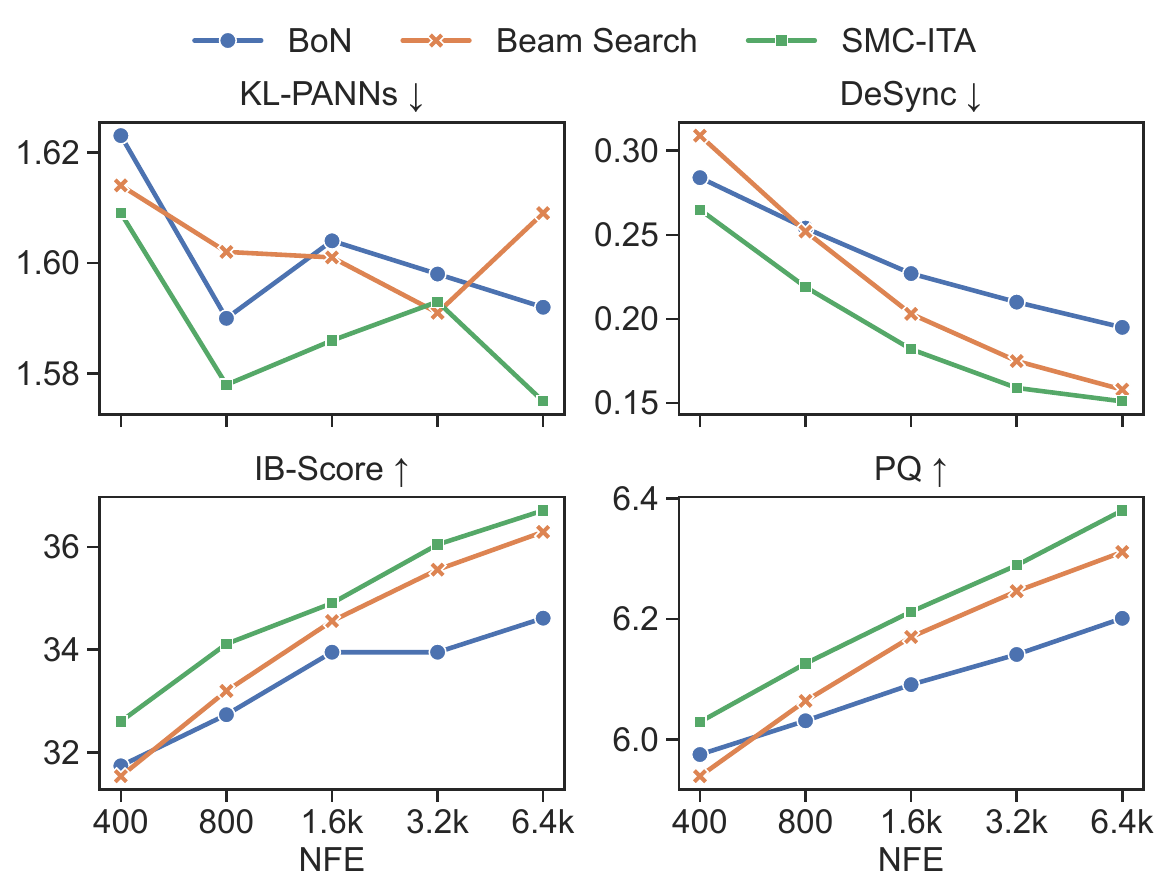}
\caption{Scaling performance. Larger NFE budgets enable search-based methods, including Beam Search and SMC-ITA, to outperform BoN. In particular, SMC-ITA shows consistent gains as the NFE budget increases.}
\label{fig:scale}
\vspace{-1mm}
\end{figure}

{\bf Scaling with different NFE budgets.}
We further evaluate the scaling behavior under different NFE budgets. As the NFE budget increases, DeSync, IB-Score, and PQ consistently improve. KL-PANNs also shows an overall improvement trend, although with some fluctuations. Both Beam Search and SMC-ITA perform search at intermediate steps, and their advantage over BoN becomes more evident as the NFE budget increases. Under the same NFE budget, SMC-ITA achieves better overall performance than Beam Search, suggesting more effective use of the sampling budget in our setting.

\section{Conclusion}
In this paper, we studied inference-time alignment for video-to-audio generation from a search perspective, rather than through training-time or architectural modifications.
We proposed SMC-ITA, which combines lookahead-based reward estimation with sequential Monte Carlo resampling to reallocate a fixed NFE budget toward more promising trajectories during flow matching sampling. 
Results show that SMC-ITA achieves a better overall alignment-quality trade-off than strong inference-time baselines across multiple evaluation dimensions.
Our analysis suggests that the gains come not only from maintaining multiple trajectories, but from making intermediate reward signals more informative for search.
In future work, developing adaptive search methods with lower rollout cost may further improve both efficiency and performance.

\bibliographystyle{IEEEtran}

\bibliography{mybib}

\end{document}